\documentclass[reprint, superscriptaddress, amsmath, amssymb, aps, prb]{revtex4-2}
\usepackage[dvipdfmx]{graphicx}% Include figure files
\usepackage{dcolumn}% Align table columns on decimal point
\usepackage{bm}% bold math
\usepackage[dvipdfmx]{hyperref}% add hypertext capabilities
\usepackage{comment}
\usepackage{color}
\usepackage{physics}
\begin{document}
\preprint{APS/123-QED}
\title{Noise-to-current ratio divergence as a fingerprint of dispersing Majorana edge modes}
\author{Leo Katayama}
\thanks{These authors contributed equally to this work.}
\affiliation{Department of Applied Physics, Nagoya University, Nagoya 464-8603, Japan}
\author{Andreas P. Schnyder}
\affiliation{Max-Planck-Institut f\"ur Festk\"orperforschung, Heisenbergstrasse 1, D-70569 Stuttgart, Germany}
\author{Yasuhiro Asano}
\affiliation{Department of Applied Physics, Hokkaido University, Sapporo 060-8628, Japan}
\author{Satoshi Ikegaya}
\thanks{These authors contributed equally to this work.}
\affiliation{Department of Applied Physics, Nagoya University, Nagoya 464-8603, Japan}
\affiliation{Institute for Advanced Research, Nagoya University, Nagoya 464-8601, Japan}
\date{\today}

%****************************************************************************************************************
\begin{abstract}
The definitive detection of Majorana modes in topological superconductors is a key issue in condensed matter physics.
Here we propose a smoking-gun experiment for the detection of one-dimensional dispersing Majorana edge modes,
based on theoretical results for multi-terminal transport in a setup consisting of two normal metal leads and a topological superconductor.
In the proposed device, the unpaired nature of the Majorana edge modes inherently leads to the absence of the charge current in the linear response regime, while the current fluctuation remains significant.
Therefore, the divergence in the noise-to-current ratio serves as unambiguous evidence for the presence of the dispersing Majorana edge modes.
We reach this conclusion analytically, without relying on any specific model of topological superconductors.
In addition, using tight-binding models of topological-insulator-based topological superconductors, we numerically verify the predicted divergent noise-to-current ratio.
We also discuss the application of our proposal to the CoSi$_2$/TiSi$_2$ heterostructure and the iron-based superconductor FeTe$_{1-x}$Se$_x$.
\end{abstract}
\maketitle
%****************************************************************************************************************

%****************************************************************************************************************
\section{Introduction}
%****************************************************************************************************************
Majorana modes in topological superconductors (TSs)~\cite{green_00,kitaev_01,wilczek_09,kane_10,zhang_11} have become a central topic in condensed matter physics,
due to their potential applications for fault-tolerant topological quantum information technologies.~\cite{ivanov_01,kitaev_03,sarma_08,fisher_11,alicea_16}.
To date, the existence of Majorana edge modes (MEMs) has been predicted in various platforms,
including semiconductor-superconductor hybrids~\cite{sato_09(1),sarma_10(1),sarma_10(2),oreg_10,kouwenhoven_12,flensberg_17,halperin_17,nichele_19,yacoby_19},
magnetic material-superconductor hybrids~\cite{beenakker_11,yazdani_13,yazdani_14,cren_17},
topological insulator-superconductor hybrids~\cite{kane_08,zhang_10,zhang_15},
Cu$_x$Bi$_2$Se$_3$~\cite{ando_11,fu_12,tanaka_12},
and iron-based superconductors~\cite{kane_13,zfang_17,zhang_16,pan_15,shin_18,gao_18,tamegai_19,sarma_21}.
However, the unambiguous verification of MEMs in these systems remains a significant challenge~\cite{microsoft_25}.
Zero-bias conductance peaks observed in normal metal--superconductor junctions~\cite{buchholtz_81,hu_94,tanaka_95,sarma_01,asano_04,law_09}
have long been regarded as a strong experimental signature of MEMs.
Nevertheless, recent studies have shown that zero-bias conductance peaks can also arise from various factors unrelated to Majorana physics%
~\cite{brouwer_12,aguado_12,sarma_17,klinovaja_20,chang_20,katsaros_21,frolov_21,ando_24}, making their definitive identification difficult.
This stalemate highlights the necessity of further experimental strategies for the unambiguous detection of MEMs.

%----------------------------------------------------------------------------
\begin{figure}[b]
\begin{center}
\includegraphics[width=0.42\textwidth]{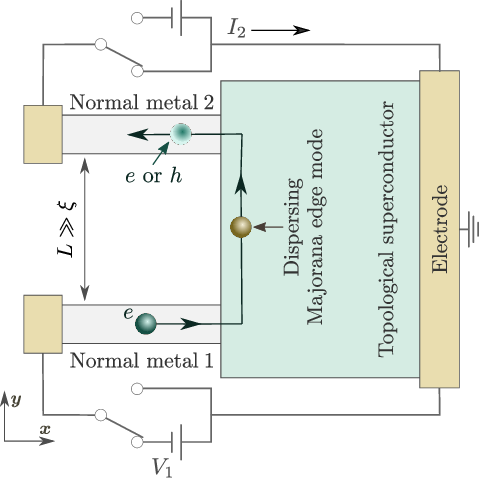}
\caption{Schematic image of the setup consisting of two NM leads and a TS.
We measure the charge current $I_{\alpha}$ in the $\alpha$-th NM due to the bias voltage $V_{\beta}$ applied to the $\beta$-th NM, where $\alpha \neq \beta$.
The $\alpha$-th NM and the TS are grounded (i.e., $V_{\alpha}=0$).
In Figure, we illustrate the case for $(\alpha,\beta)=(2,1)$,
where a dispersing MEM mediates inter-lead scattering from an electron in the first NM to either an electron or a hole in the second NM.
}
\label{fig:figure1}
\end{center}
\end{figure}
%----------------------------------------------------------------------------
In this Letter, we propose a versatile experiment that provides a smoking-gun signature of the one-dimensional dispersing MEMs.
The one-dimensional dispersing MEMs are particularly important because they can be harnessed in the braiding operations of topological computing~\cite{beenakker_19,beenakker_19(2),beenakker_20}.
To detect the dispersing MEMs, we measure the nonlocal charge transport~\cite{law_09,akhmerov_09,kane_09(2),beenakker_10,beri_12,beenakker_15(2),ikegaya_19}
in a multi-terminal setup as shown in Fig.~\ref{fig:figure1}, where two normal metal (NM) leads are attached to one edge of a TS.
When the distance between the NM leads ($L$) is sufficiently larger than the superconducting coherence length ($\xi$),
inter-lead scattering mediated by evanescent modes of conventional Bogoliubov quasiparticles vanishes completely~\cite{deutscher_00}.
Thus, the nonlocal charge current at bias voltages below the superconducting gap is carried exclusively by the MEM propagating along the edge.
Importantly, in usual Bogoliubov quasiparticles, electron-like and hole-like modes exist in \emph{pairs} at zero energy due to particle-hole symmetry.
In contrast, the MEM, where particle and antiparticle modes are equivalent, can exist individually.
We show that this \emph{unpaired} nature of the MEM essentially leads to the absence of the charge currents in the linear response limit, while the current fluctuation remains significant.
As a result, the divergence in the noise-to-current ratio at zero bias voltage serves as a smoking-gun signature of the dispersing MEM in the proposed setup.
We reach this conclusion without relying on any specific model of the TS.
Thus, we expect that the proposed experiment is useful to detect any one-dimensional dispersing MEMs,
including Majorana \emph{hinge} modes in three-dimensional second-order TSs~\cite{balents_15,hughes_17,fang_17,brouwer_17,khalaf_18}. 
To validate our theory more robustly, we also perform numerical calculations
for specific tight-binding models of the superconducting topological insulator thin films~\cite{zhang_15,sarma_21}.
Finally, we discuss potential applications of our proposal to the CoSi$_2$/TiSi$_2$ heterostructure~\cite{lin_21,lin_23}
and the iron-based superconductor FeTe$_{1-x}$Se$_x$~\cite{sarma_19,kim_21,zhang_24,burch_19}.

%****************************************************************************************************************
\section{noise-to-current ratio} \label{sec:2}
%****************************************************************************************************************
Let us consider a two-dimensional hybrid system as shown in Fig.~\ref{fig:figure1}, where two NM leads are attached to an edge of a TS.
We calculate the charge current at zero temperature within the Blonder--Tinkham--Klapwijk (BTK) formalism~\cite{deutscher_00,klapwijk_82}:
\begin{align}
\begin{split}
&\bar{I}_{\alpha} = \frac{1}{e}\int^{eV_{\beta}}_{0} G_{\alpha \beta}(E) dE,\\
&G_{\alpha\beta}(E)=-\frac{e^2}{h}\mathrm{Tr}\left[ \hat{T}^{e}_{\alpha\beta}(E) - \hat{T}^{h}_{\alpha\beta}(E)\right],\\
&\hat{T}^{\gamma}_{\alpha\beta}(E)= \{\hat{s}^{\gamma e}_{\alpha\beta}(E) \}^{\dagger}\hat{s}^{\gamma e}_{\alpha\beta}(E),
\end{split} \label{eq:diff_cond}
\end{align}
where we assume $\alpha \neq \beta$.
$\bar{I}_{\alpha}$ represents the time-averaged charge current in the $\alpha$-th NM due to the bias voltage $V_{\beta}$ applied to the electrode connected to the $\beta$-th NM,
where the electrodes attached to the $\alpha$-th NM and the TS are grounded (i.e., $V_{\alpha}=0$).
We denote the number of propagating channels in the $\alpha$-th NM by $N_{\alpha}$.
The $N_{\alpha} \times N_{\beta}$ matrices of $\hat{s}^{ee}_{\alpha\beta}$ and $\hat{s}^{he}_{\alpha\beta}$ ($\hat{s}^{eh}_{\alpha\beta}$ and $\hat{s}^{hh}_{\alpha\beta}$) contain
the scattering coefficients from an electron (a hole) in the $\beta$-th NM to an electron and a hole in the $\alpha$-th NM at energy $E$, respectively.
In addition, we calculate the zero-frequency noise power based on the BTK formalism~\cite{beenakker_94,datta_96},
\begin{align}
C_{\alpha} = \int^{\infty}_{-\infty} \overline{\delta I_{\alpha}(0) \delta I_{\alpha}(\tau)} d\tau =  \frac{1}{e} \int^{eV_{\beta}}_{0} P_{\alpha\beta}(E) dE,
\end{align}
with
\begin{align}
\begin{split}
P_{\beta \alpha}(E) =& \frac{e^3}{h}\mathrm{Tr} \left[ \hat{T}^{e}_{\alpha\beta}(E) + \hat{T}^{h}_{\alpha\beta}(E) \right]\\
-&\frac{e^3}{h}\mathrm{Tr}\left[
\{ \hat{T}^{e}_{\alpha\beta}(E) - \hat{T}^{h}_{\alpha\beta}(E) \}^2 \right] ,
\label{eq:noise_power}
\end{split}
\end{align}
where $\delta I_{\alpha}(\tau) = I_{\alpha}(\tau) - \bar{I}_{\alpha}$ represents the deviation of the current at time $\tau$ from the time-averaged current.
In the BTK formalism, we assume that the currents flowing towards $x=+\infty$ ($x=-\infty$) are absorbed into the ideal electrode connected to the TS (NM),
while these electrodes are not explicitly included in the Hamiltonian.
The BTK formalism is quantitatively valid for bias voltages well below the superconducting gap.

In the following, we focus on the linear response limit $V_{\beta} \rightarrow 0$,
where the time-averaged current, $\bar{I}_{\alpha}=G_{\alpha\beta}(0)V_{\beta}$,
and the zero-frequency noise power, $C_{\alpha} = P_{\alpha \beta}(0) V_{\beta}$, are calculated using only the inter-lead scattering matrix at zero energy:
\begin{align}
\hat{s}_{\alpha\beta}(E=0)= \left[ \begin{array}{cc}
\hat{s}^{ee}_{\alpha\beta} & \hat{s}^{eh}_{\alpha\beta} \\
\hat{s}^{he}_{\alpha\beta} & \hat{s}^{hh}_{\alpha\beta} \end{array} \right], \qquad (\alpha \neq \beta),
\end{align}
where we use the notation $\hat{s}^{\gamma\gamma^{\prime}}_{\alpha\beta}(E=0)=\hat{s}^{\gamma\gamma^{\prime}}_{\alpha\beta}$, for simplicity.
Due to particle-hole symmetry of the superconducting system, the scattering matrix satisfies~\cite{beenakker_15}
\begin{align}
\begin{split}
\hat{\Xi}_{\alpha} \hat{s}_{\alpha\beta}(E) \hat{\Xi}_{\beta} = \hat{s}_{\alpha\beta}(-E),\quad
\hat{\Xi}_{\alpha} = \left[ \begin{array}{cc} \hat{O}_{\alpha\alpha} & \hat{I}_{\alpha} \\ \hat{I}_{\alpha} & \hat{O}_{\alpha\alpha} \end{array} \right] \mathcal{K},
\end{split}
\end{align}
where $\hat{O}_{\alpha\beta}$ is the $N_{\alpha} \times N_{\beta}$ null matrix,
$\hat{I}_{\alpha}$ denotes the $N_{\alpha} \times N_{\alpha}$ identity matrix,
and $\mathcal{K}$ represents the complex conjugation operator.
Therefore, we obtain
\begin{align}
\hat{s}^{eh}_{\alpha\beta} = (\hat{s}^{he}_{\alpha\beta})^{\ast}, \quad
\hat{s}^{hh}_{\alpha\beta} = (\hat{s}^{ee}_{\alpha\beta})^{\ast}.
\end{align}
We rewrite the scattering matrix in a Majorana representation, transforming $\hat{s}_{\alpha\beta}(E=0)$ into a real matrix~\cite{beenakker_15}:
\begin{align}
\hat{\Gamma}_{\alpha} \hat{s}_{\alpha\beta}(E=0) \hat{\Gamma}_{\beta}^{\dagger} = \hat{s}^M_{\alpha\beta},\quad
\hat{\Gamma}_{\alpha} = \frac{1}{\sqrt{2}} \left[ \begin{array}{cc} \hat{I}_{\alpha} &  \hat{I}_{\alpha} \\ i \hat{I}_{\alpha} & -i \hat{I}_{\alpha} \end{array} \right],
\end{align}
where $(\hat{s}^M_{\alpha\beta})^{\ast}=\hat{s}^M_{\alpha\beta}$.
Using singular value decomposition, we represent the scattering matrix as
\begin{align}
\hat{s}^M_{\alpha\beta}=\hat{\mathcal{U}}_{\alpha} \hat{\Lambda}_{\alpha\beta} \hat{\mathcal{V}}_{\beta},
\end{align}
where $\hat{\Lambda}_{\alpha\beta}$ is the $2N_{\alpha} \times 2N_{\beta}$ real diagonal matrix,
$\hat{\mathcal{U}}_{\alpha}$ ($\hat{\mathcal{V}}_{\beta}$) is the real orthogonal matrix satisfying
$\hat{\mathcal{U}}_{\alpha}\hat{\mathcal{U}}_{\alpha}^{\mathrm{T}}=\hat{I}_{2\alpha}$ ($\hat{\mathcal{V}}_{\beta}\hat{\mathcal{V}}_{\beta}^{\mathrm{T}}=\hat{I}_{2\beta}$).

Here we assume $L \gg \xi$, where $L$ is the distance between the two NM leads and $\xi$ is the superconducting coherence length.
In this limit, the inter-lead scattering mediated by evanescent modes of the Bogoliubov quasiparticles vanishes completely~\cite{deutscher_00}.
In addition, we assume that only \emph{one} dispersing MEM mediates the inter-lead scattering from the $\beta$-th NM to the $\alpha$-th NM, as shown in Fig.~\ref{fig:figure1}.
These assumptions lead to~\cite{beenakker_15}
\begin{align}
\mathrm{rank}(\hat{\Lambda}_{\alpha\beta}) = 1.
\label{eq:rank_s}
\end{align}
Note that for conventional Bogoiubov quasiparticles, the electron-like and hole-like modes exist in \emph{pairs} at zero energy due to particle-hole symmetry,
which results in an \emph{even} rank for $\hat{\Lambda}_{\alpha\beta}$.
Therefore, the relation of $\mathrm{rank}(\hat{\Lambda}_{\alpha\beta}) = 1$ is a clear manifestation of the \emph{unpaired} nature of the MEM.
For a TS belonging to class D of the Altland-Zirnbauer symmetry class~\cite{schnyder_08},
we specifically assume a $\mathbb{Z}$ topological invariant of $\nu=\pm1$, which leads to the appearance of a \emph{single} chiral Majorana edge mode (CMEM).
The direction of motion of the CMEM is associated with the sign of $\nu$.
When the CMEM moves in the direction from the $\beta$-th NM to the $\alpha$-th NM,
we have $\mathrm{rank}(\hat{\Lambda}_{\alpha\beta}) = 1$, while $\mathrm{rank}(\hat{\Lambda}_{\beta\alpha}) = 0$.
When the TS belongs to class DIII~\cite{schnyder_08} and has a $\mathbb{Z}_2$ topological invariant of $\nu=1$, we obtain a \emph{single} pair of helical Majorana edge modes (HMEM).
In this case, we have $\mathrm{rank}(\hat{\Lambda}_{\alpha\beta}) = \mathrm{rank}(\hat{\Lambda}_{\beta\alpha}) = 1$.

Using the representation of
\begin{align}
\begin{split}
&\hat{\Lambda}_{\alpha\beta}=\left[ \begin{array}{cc}
\hat{\lambda}_{\alpha\beta} & \hat{O}_{\alpha\beta} \\
\hat{O}_{\alpha\beta}& \hat{O}_{\alpha\beta} \end{array} \right], \\
&\hat{\lambda}_{\alpha\beta}=\mathrm{diag}[\lambda_{\alpha\beta},0,\cdots,0],\\
&\hat{\mathcal{U}}_{\alpha}=\left[ \begin{array}{cc}
\hat{u}_{++} & \hat{u}_{+-} \\
\hat{u}_{-+} & u_{--} \end{array} \right], \quad
\hat{\mathcal{V}}_{\alpha}=\left[ \begin{array}{cc}
\hat{v}_{++} & \hat{v}_{+-} \\
\hat{v}_{-+} & v_{--} \end{array} \right],
\end{split}
\end{align}
where $\hat{\lambda}_{\alpha\beta}$ is the $N_{\alpha} \times N_{\beta}$ diagonal matrix with a single nonzero element $\lambda_{\alpha\beta}$,
and $\hat{u}_{ss^{\prime}}$ ($\hat{v}_{ss^{\prime}}$) is the $N_{\alpha} \times N_{\alpha}$ ($N_{\beta} \times N_{\beta}$) matrix satisfying
\begin{align}
\begin{split}
&\hat{u}_{++}\hat{u}_{++}^{\mathrm{T}}+\hat{u}_{+-}\hat{u}_{+-}^{\mathrm{T}} = \hat{I}_{\alpha},\\
&\hat{v}_{++}\hat{v}_{++}^{\mathrm{T}}+\hat{v}_{+-}\hat{v}_{+-}^{\mathrm{T}} = \hat{I}_{\beta},
\end{split}
\end{align}
we rewrite the scattering matrix as
\begin{align}
\hat{s}^M_{\alpha\beta}=\left[ \begin{array}{cc}
\hat{u}_{++}\hat{\lambda}_{\alpha\beta}\hat{v}_{++}& \hat{u}_{++}\hat{\lambda}_{\alpha\beta}\hat{v}_{+-} \\
\hat{u}_{-+}\hat{\lambda}_{\alpha\beta}\hat{v}_{++} & \hat{u}_{-+}\hat{\lambda}_{\alpha\beta}\hat{v}_{+-} \end{array} \right].
\end{align}
Applying the unitary transformation,
\begin{align}
\hat{\Gamma}_{\alpha}^{\dagger} \hat{s}^M_{\alpha\beta} \hat{\Gamma}_{\beta} = \hat{s}_{\alpha\beta}(E=0),
\end{align}
we obtain the scattering matrix in the original basis as
\begin{align}
\begin{split}
&\hat{s}^{ee}_{\alpha\beta} = \frac{1}{2} (\hat{u}_{++} - i \hat{u}_{-+}) \hat{\lambda}_{\alpha\beta} (\hat{v}_{++} + i \hat{v}_{+-}),\\
&\hat{s}^{he}_{\alpha\beta} = \frac{1}{2} (\hat{u}_{++} + i \hat{u}_{-+}) \hat{\lambda}_{\alpha\beta} (\hat{v}_{++} + i \hat{v}_{+-}).
\end{split}
\end{align}
Using the equations,
\begin{align}
\begin{split}
&\hat{\lambda}^{\mathrm{T}}_{\alpha\beta}(\hat{u}_{++}^{\mathrm{T}} \pm i \hat{u}_{-+}^{\mathrm{T}}) (\hat{u}_{++} \mp i \hat{u}_{-+})\hat{\lambda}_{\alpha\beta}
= \hat{\lambda}^{\mathrm{T}}_{\alpha\beta} \hat{\lambda}_{\alpha\beta},\\
&\hat{\lambda}_{\alpha\beta}(\hat{v}_{++} \pm i \hat{v}_{+-}) (\hat{v}_{++}^{\mathrm{T}} \mp i \hat{v}_{+-}^{\mathrm{T}}) \hat{\lambda}^{\mathrm{T}}_{\alpha\beta}
=  \hat{\lambda}_{\alpha\beta} \hat{\lambda}^{\mathrm{T}}_{\alpha\beta},
\end{split}
\end{align}
we obtain
\begin{align}
\begin{split}
&\mathrm{Tr}[\hat{T}^{\gamma}_{\alpha\beta}(E=0)] = \frac{\lambda_{\alpha\beta}^2}{4},\\
&\mathrm{Tr}[\hat{T}^{\gamma}_{\alpha\beta}(E=0)\hat{T}^{\gamma^{\prime}}_{\alpha\beta}(E=0)] = \frac{\lambda_{\alpha\beta}^4}{16}.
\end{split}
\end{align}
Substituting this into Eq.~(\ref{eq:diff_cond}) and Eq.~(\ref{eq:noise_power}), we eventually find
\begin{align}
\begin{split}
&G_{\alpha \beta}(0)  = 0,\\
&P_{\alpha \beta}(0)  = \frac{e^3 \lambda_{\alpha \beta}^2}{2h} \neq 0,
\label{eq:linear_res}
\end{split}
\end{align}
meaning that in the linear response regime $V_{\beta}\rightarrow 0$,
the time-averaged current $\bar{I}_{\alpha}=G_{\alpha\beta}(0)V_{\beta}$ vanishes while the noise $C_{\alpha} = P_{\alpha \beta}(0) V_{\beta}$ remains nonzero.
This characteristic is highlighted by the divergence in the noise-to-current ratio at zero-bias voltage:
\begin{align}
\lim_{V_{\beta} \rightarrow 0} F_{\alpha}(eV_{\beta}) = 
\lim_{V_{\beta} \rightarrow 0} \frac{C_{\alpha} (eV_{\beta}) } {e |\bar{I}_{\alpha} (eV_{\beta})|}  = \infty.
\label{eq:ns_divergence}
\end{align}
In the absence of dispersing MEMs, we obtain $\lambda_{\alpha \beta}=0$, which leads to $G_{\beta \alpha}(0)  = P_{\beta \alpha}(0) = 0$.
When Bogoliubov quasiparticle modes other than the unpaired MEM mediate inter-lead scatterings, we obtain $G_{\beta \alpha}(0) \neq 0$.
Therefore, the transport properties presented in Eq.~(\ref{eq:linear_res}) and Eq.~(\ref{eq:ns_divergence}) serve as a clear manifestation of the propagating and unpaired nature of the dispersing MEM.
The results above is obtained without considering any specific model of TSs.
Thus, we expect that the proposed experiment is applicable to observe various one-dimensional dispersing MEMs,
including CMEMs/HMEMs in \emph{general} two-dimensional TSs with $|\nu|=1$, as well as Majorana \emph{hinge} states in three-dimensional second-order TSs.

\vspace{20pt}
%****************************************************************************************************************
\section{Numerical Confirmation}
%****************************************************************************************************************
%----------------------------------------------------------------------------
\begin{figure}[b]
\begin{center}
\includegraphics[width=0.45\textwidth]{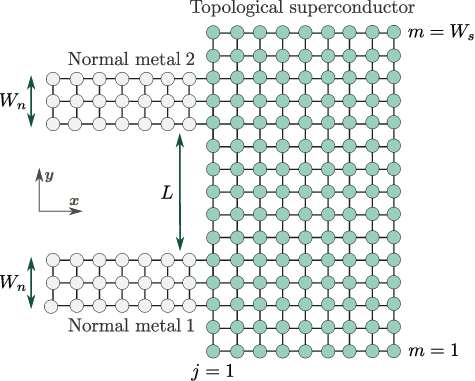}
\caption{Proposed setup on a two-dimensional tight-binding model.}
\label{fig:figure2}
\end{center}
\end{figure}
%----------------------------------------------------------------------------
We numerically reproduce Eq.~(\ref{eq:linear_res}) and Eq.~(\ref{eq:ns_divergence}) using explicit models of TSs.
We describe the present setup using a two-dimensional tight-binding model on a square lattice, as shown in Fig.~\ref{fig:figure2}.
A lattice site is indicated by a vector $\boldsymbol{r}=j \boldsymbol{x} + m \boldsymbol{y}$, where $\boldsymbol{x}$ and $\boldsymbol{y}$ are unit vectors with length $a_0=1$.
The TS occupies $1 \leq j \leq \infty$ and $ 1 \leq m \leq W_s$.
The $\alpha$-th NM is placed on $- \infty \leq j \leq 0$ and $m_{\alpha} < m \leq M_{\alpha}$, where $M_{\alpha}=m_{\alpha}+W_n$ and $m_2 = M_1+L$.
We denote the width of the NM by $W_n$ and the distance between the first NM and the second NM by $L$.
In the $y$ direction, we apply an open boundary condition.

We discuss the topological superconductivity realized in superconducting topological insulator thin films~\cite{zhang_15,sarma_21}.
As will be discussed in detail later, the model we use can describe TSs belonging to both class D and class DIII by tuning the parameters~\cite{zhang_15,sarma_21}.
The model is given by the Bogoliubov--de Gennes Hamiltonian,
\begin{align}
\begin{split}
&H_s = \frac{1}{2} \sum_{\boldsymbol{k}}\left[C_{\boldsymbol{k}}^{\dagger},C_{-\boldsymbol{k}}^{\mathrm{T}}\right] 
\left[\begin{array}{cc} h_{\boldsymbol{k}} & \Delta_{\boldsymbol{k}}\\
\Delta_{\boldsymbol{k}}^{\dagger} & -h_{-\boldsymbol{k}}^{\ast} \end{array}\right]
\left[\begin{array}{cc} C_{\boldsymbol{k}} \\ C_{-\boldsymbol{k}}^{\ast} \end{array} \right],\\
&C_{\boldsymbol{k}}=\left[c_{\boldsymbol{k}\uparrow t}, c_{\boldsymbol{k}\downarrow t}, c_{\boldsymbol{k}\uparrow b}, c_{\boldsymbol{k}\downarrow b} \right]^{\mathrm{T}},\\
&h_{\boldsymbol{k}} = h^{\mathrm{Dirac}}_{\boldsymbol{k}} + V,\\
&h^{\mathrm{Dirac}}_{\boldsymbol{k}} = v  \rho_z \otimes  (\sin k_y s_x - \sin k_x s_y) + w_{\boldsymbol{k}}\rho_x,\\
&w_{\boldsymbol{k}} = w(2-\cos k_x - \cos k_y),\\
&V = t \rho_x- (\mu_0 + \mu_{\delta} \rho_z) + (m_0 + m_{\delta} \rho_z) \otimes s_z,\\
&\Delta_{\boldsymbol{k}} = \left\{\Delta_0 + \Delta_1 (\cos k_x + \cos k_y) \right\} (i s_y),
\end{split} \label{eq:tb_ham}
\end{align}
where we assume that the low-energy physics consists only of Dirac surface states at the top and bottom layers of the topological insulator thin film.
$c_{\boldsymbol{k}s t}$ ($c_{\boldsymbol{k}s b}$) is an annihilation operator of a Dirac electron at the top (bottom) layer with momentum $\boldsymbol{k}$ and spin $s=\uparrow, \downarrow$.
The Pauli matrices $\rho_{\nu}$ and $s_{\nu}$ for $\nu=x,y,z$ denote the layer and spin degrees of freedom, respectively.
The Hamiltonian $h^{\mathrm{Dirac}}_{\boldsymbol{k}}$ describes the Dirac surface states.
The second term $w_{\boldsymbol{k}}$, which is similar to the Wilson mass term~\cite{wilson_74,sun_17},
opens gaps at the spurious Dirac points located at $(k_x,k_y)=(\pm \pi, \pm \pi)$, while leaving the Dirac point at $(k_x,k_y)=(0,0)$ gapless.
The coupling between the top and bottom Dirac surface states is denoted by $t$.
The chemical potential at the top (bottom) layer is given by $\mu_0+\mu_{\delta}$ ($\mu_0-\mu_{\delta}$),
where the potential difference between the top and bottom layers can be induced by an applied electric field along the out-of-plane direction~\cite{sarma_21}.
We also consider the exchange potential along the $z$ axis, which can be induced by
the intrinsic ferromagnetic ordering~\cite{johnson_18}, magnetic doping~\cite{shen_10}, or magnetic proximity effects~\cite{moodera_13},
where the exchange potential at the top (bottom) layer is represented by $m_0+m_{\delta}$ ($m_0-m_{\delta}$).
We assume that the extended $s$-wave pair potential in the bulk provides a self-proximity effect on the Dirac surface states~\cite{shin_18},
where the effective pair potential acting on the Dirac surface states is denoted by $\Delta_{\boldsymbol{k}}$.
For the numerical calculations, we rewrite $H_s$ in a real space basis by applying a Fourier transformation: $c_{\boldsymbol{k}s t (b)} \rightarrow c_{\boldsymbol{r}s t (b)}$.
The $\alpha$-th NM is described by
\begin{align}
\begin{split}
H_{N_{\alpha}}=&-t_n \sum_{\boldsymbol{r}\in N_{\alpha}}\sum_{s=\uparrow,\downarrow}\sum_{\boldsymbol{\nu}=\boldsymbol{x},\boldsymbol{y}}
(d_{\boldsymbol{r}+\boldsymbol{\nu}s}^{\dagger}d_{\boldsymbol{r}s} + \mathrm{H.c.})\\
&+(4t_n-\mu_n) \sum_{\boldsymbol{r}\in N_{\alpha}}\sum_{s=\uparrow,\downarrow}d_{\boldsymbol{r}s}^{\dagger}d_{\boldsymbol{r}s},
\end{split}
\end{align}
where $d_{\boldsymbol{r}s}$ annihilates an electron at a site $\boldsymbol{r}$ with spin $s$,
and $\sum_{\boldsymbol{r}\in N_{\alpha}}$ represents the summation over the lattice sites in the $\alpha$-th NM.
We describe the junction interface between the TS and the $\alpha$-th NM by
\begin{align}
H_{T_{\alpha}} =&-t_{\alpha t}\sum_{m>m_{\alpha}}^{M_{\alpha}}\sum_{s=\uparrow,\downarrow}(c_{j=1,m,s,t}^{\dagger}d_{j=0,m,s}+\mathrm{H.c.}) \nonumber\\
&-t_{\alpha b}\sum_{m>m_{\alpha}}^{M_{\alpha}}\sum_{s=\uparrow,\downarrow}(c_{1,m,s,b}^{\dagger}d_{0,m,s}+\mathrm{H.c.}),
\end{align}
where we assume that the coupling is independent of the spins, while it depends on the leads and layers.
In the following calculations, we fix the parameters as $W_s=150$, $W_n=10$, $L=110$, $m_1=10$, and $\mu_N=t_n$.
To highlight the robustness of our conclusion, we assume the irregular couplings at the interfaces:
$(t_{1t},t_{1b},t_{2t},t_{2b})=(0.5t_n,0.2t_n,0.3t_n,0.4t_n)$, satisfying $t_{1t} \neq t_{1b} \neq t_{2t} \neq t_{2b}$.
The scattering coefficients in Eq.~(\ref{eq:diff_cond}) and Eq.~(\ref{eq:noise_power}) are numerically computed using recursive Green's function techniques~\cite{fisher_81,ando_91}.

\subsection{Class D}
\begin{figure}[t]
\begin{center}
\includegraphics[width=0.5\textwidth]{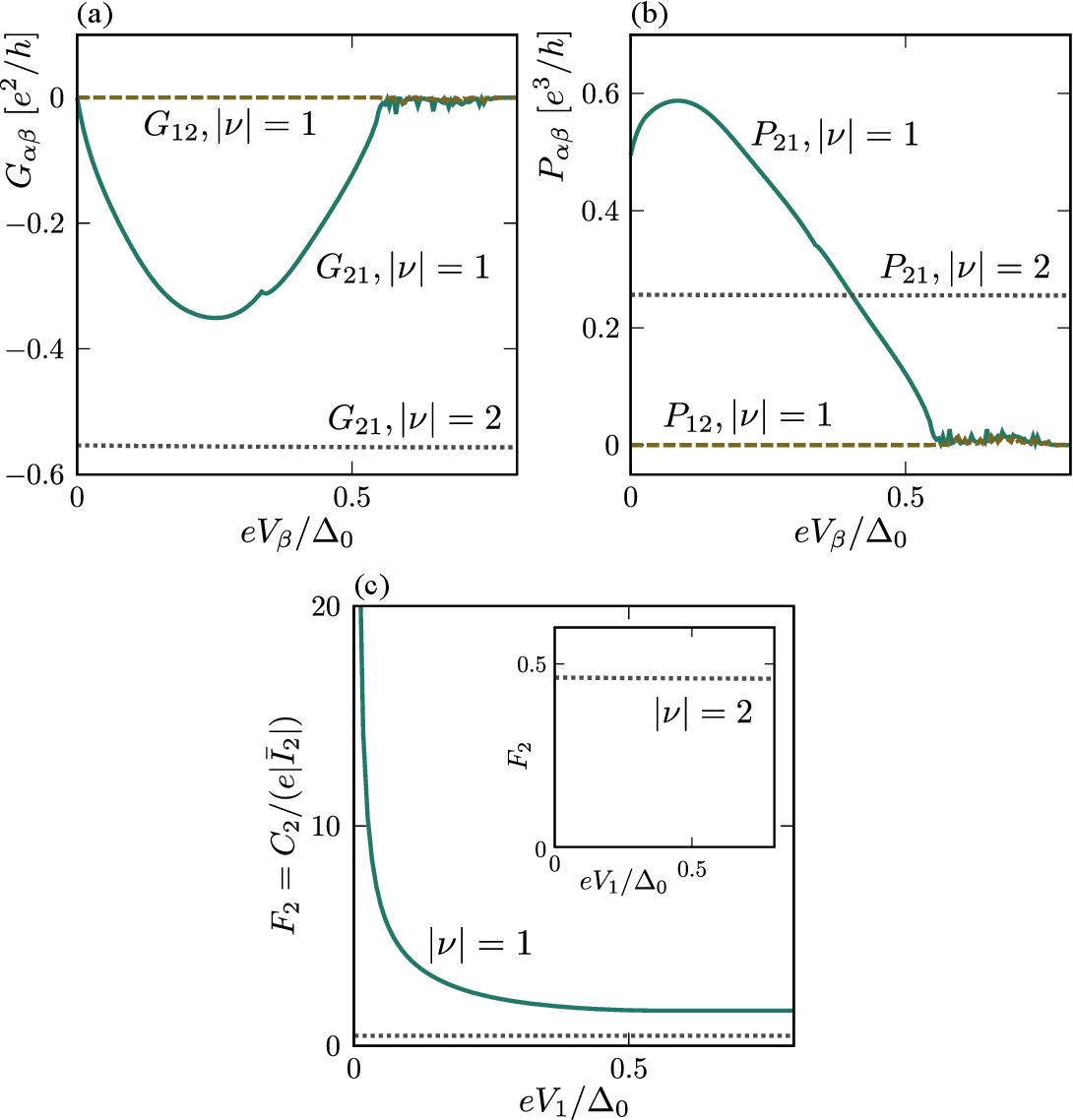}
\caption{Multi-terminal transport properties of the class D TS.
(a) Differential conductance $G_{\alpha \beta}$, (b) $P_{\alpha \beta}=dC_{\alpha}/dV_{\beta}$
and (c) noise-to-current ratio $F_2$ as a function of the bias voltage.
The solid (dotted) lines show the results for $|\nu|=1$ ($|\nu|=2$) with $(\alpha, \beta)=(2,1)$.
The dashed lines represents the results for $|\nu|=1$ with $(\alpha, \beta)=(1,2)$.
The results show perfect agreement with the analytical predictions presented in Sec.~ \ref{sec:2}.
}
\label{fig:figure3}
\end{center}
\end{figure}

First, we consider a TS belonging to class D, where we set the parameters as
$v=t_n$, $w=2t_n$, $m_0=t_n$, $m_{\delta}=0.5t_n$, and $t=\mu_{\delta}=\Delta_1=0$.
We assume the uniform chemical potential (i.e., $\mu_{\delta}=0$) and the simple $s$-wave pair potential (i.e., $\Delta_1=0$), for simplicity.
Note that a very similar model was proposed in Ref.~[\onlinecite{zhang_15}];
the pair potential  in Ref.~[\onlinecite{zhang_15}] has the layer dependence, while the exchange potential in our model has the layer dependence.
For $m_0+m_{\delta}>\mu_0$, as shown in detail in the Supplemental Material (SM)~\cite{sm},
the low-energy effective Hamiltonian for the superconducting segment is given by
\begin{align}
\begin{split}
&H^{\mathrm{eff}}_{\boldsymbol{k}}=\left[ \begin{array}{cc}
h^{\mathrm{QAH}}_{\boldsymbol{k}} & \Delta^{\mathrm{eff}} (i \sigma_y)
\\ -\Delta^{\mathrm{eff}} (i \sigma_y) & -(h^{\mathrm{QAH}}_{-\boldsymbol{k}})^{\ast} \end{array} \right],\\
&h^{\mathrm{QAH}}_{\boldsymbol{k}}=M_{\boldsymbol{k}} \sigma_z -\mu_0 + v(\sin k_y \sigma_x + \sin k_x \sigma_y ),\\
&M_{\boldsymbol{k}}=\sqrt{w_{\boldsymbol{k}}^2+m_{\delta}^2}-m_0, \quad
\Delta^{\mathrm{eff}}=\frac{m_{\delta} \Delta}{\sqrt{w_{\boldsymbol{k}}^2+m_{\delta}^2}},
\end{split}
\end{align}
where $\boldsymbol{\sigma}=(\sigma_x,\sigma_y,\sigma_z)$ are the three Pauli matrices.
The effective Hamiltonian $H^{\mathrm{eff}}_{\boldsymbol{k}}$ is equivalent to the Hamiltonian of a superconducting quantum anomalous Hall insulator~\cite{zhang_10}.
According to Ref.~[\onlinecite{zhang_10}], the Hamiltonian $H^{\mathrm{eff}}_{\boldsymbol{k}}$ is characterized by
a $\mathbb{Z}$ topological invariant, with a value of $\nu=-2$ for $\sqrt{\mu_0^2+\Delta_0^2}<m_0-m_{\delta}$,
and $\nu=-1$ for $m_0-m_{\delta}< \sqrt{\mu_0^2+\Delta_0^2}<\sqrt{(4w)^2+m_{\delta}^2}-m_0$.
The system with $|\nu|=2$ is topologically equivalent to the normal quantum anomalous Hall insulator, and hosts the residue of the normal chiral edge modes.
The system with $|\nu|=1$ hosts single unpaired CMEMs moving in the direction from the first to the second NMs.
The exact topological phase boundary is numerically computed in the SM~\cite{sm}.
In the following, we choose $(\mu_0,\Delta_0)=(0.15t_n,0.1t_n)$ and $(\mu_0,\Delta_0)=(0.75t_n,0.5t_n)$ to describe the topological phases with $|\nu|=2$ and $|\nu|=1$, respectively.
In Fig.~\ref{fig:figure3}(a) and Fig.~\ref{fig:figure3}(b), we show $G_{\alpha \beta}$ and $P_{\alpha \beta}$ as a function of the bias voltage, respectively.
The solid lines show $G_{21}$ and $P_{21}$ with $|\nu|=1$.
The values of $G_{21}$ and $P_{21}$ for finite voltages can depend on the details of the parameters.
Nevertheless, in perfect agreement with Eq.~(\ref{eq:linear_res}) and Eq.~(\ref{eq:ns_divergence}), we find $G_{21}=0$ and $P_{21}\neq0$ at zero bias voltage.
The dashed lines show $G_{12}$ and $P_{12}$ with $|\nu|=1$.
Since the CMEM does not mediate inter-lead scatterings from the second to the first NMs, we obtain $G_{12}=P_{12}=0$ for bias voltages below the superconducting gap.
The dotted lines show the results with $|\nu|=2$.
Both $G_{21}$ and $P_{21}$ are finite due to the presence of chiral edge modes.
However, since these states are not unpaired CMEMs, we find $G_{21}\neq0$ even at zero bias voltage.
This allows us to distinguish the unpaired CMEMs from other chiral edge modes.
In Fig.~\ref{fig:figure3}(c), we show the noise-to-current ratio $F_2 = C_2/(e\bar{I}_2)$ as a function of the bias voltage.
For $|\nu|=2$, we obtain $F_2 \approx 0.5$ independent of the bias voltage.
For $|\nu|=1$, we find the divergence of the noise-to-current ratio at zero bias voltage, which is a definitive signature of the CMEM.
%----------------------------------------------------------------------------
%----------------------------------------------------------------------------

\subsection{Class DIII}
\begin{figure}[t]
\begin{center}
\includegraphics[width=0.5\textwidth]{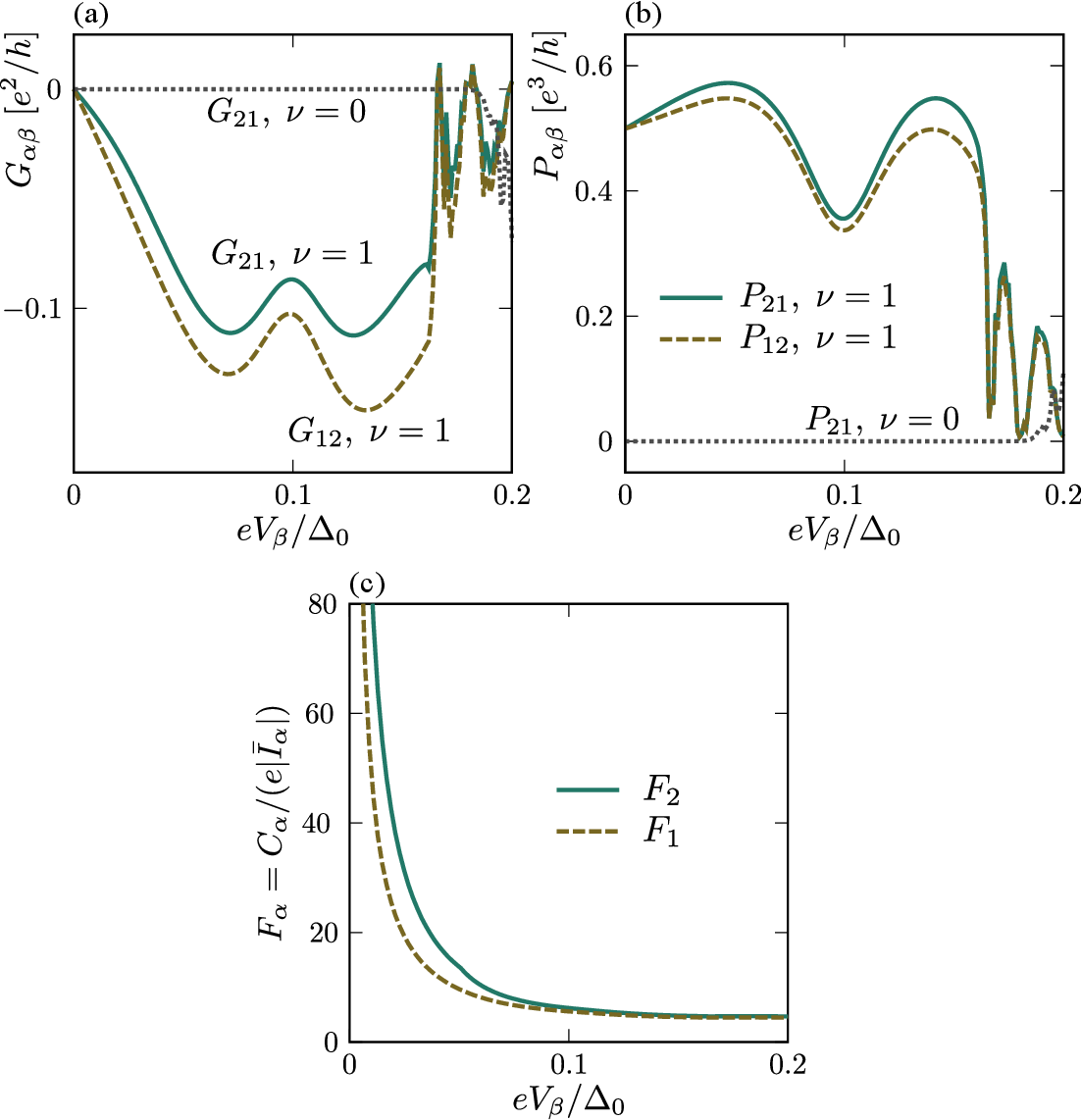}
\caption{Multi-terminal transport properties of the class DIII TS.
(a) Differential conductance $G_{\alpha \beta}$, (b) $P_{\alpha \beta}=dC_{\alpha}/dV_{\beta}$
and (c) noise-to-current ratio $F_{\alpha}$ as a function of the bias voltage.
The solid (dotted) lines show the results for $\nu=1$ ($\nu=0$) with $(\alpha, \beta)=(2,1)$.
The dashed lines represents the results for $\nu=1$ with $(\alpha, \beta)=(1,2)$.
The results show perfect agreement with the analytical predictions presented in Sec.~ \ref{sec:2}.}
\label{fig:figure4}
\end{center}
\end{figure}

Next, we consider a TS belonging to class DIII, where we set the parameters as
$v=t_n$, $w=0.5t_n$, $t=0.5t_n$, $\mu_0=t_n$, and $\mu_{\delta}=0.5t_n$.
To preserve time-reversal symmetry, we remove the exchange potential, $m_0=m_{\delta}=0$.
This model has been already proposed in Ref.~[\onlinecite{sarma_21}].
For $\sqrt{\mu_{\delta}^2+t^2}<\mu_0<\sqrt{\mu_{\delta}^2+(t+2w)^2}$, as discussed in detail in the SM~\cite{sm}, we have two Fermi surfaces surrounding the $\Gamma$ point of the Brillouin zone.
The Fermi surfaces are located at momenta $\boldsymbol{k}^{\pm}_F$ satisfying $\varepsilon_{\boldsymbol{k}^{\pm}_F,\pm}=0$ with
\begin{align}
\begin{split}
&\varepsilon_{\boldsymbol{k},\pm}=\sqrt{(\xi_{\boldsymbol{k}}\mp \mu_{\delta})^2+(w_{\boldsymbol{k}} + t)^2}-\mu_0,\\
&\xi_{\boldsymbol{k}}=v\sqrt{\sin^2 k_x + \sin^2 k_y}.
\end{split}
\end{align}
The extended $s$-wave pair potential has zeros at $\boldsymbol{k}_{\Delta}=(k_{\Delta,x},k_{\Delta,y})$ satisfying
\begin{align}
\Delta_0 + \Delta_1 (\cos k_{\Delta,x} + \cos k_{\Delta,y}) = 0.
\end{align}
When the condition,
\begin{align}
|\boldsymbol{k}_F^-| < |\boldsymbol{k}_{\Delta}| < |\boldsymbol{k}_F^+|,
\end{align}
is satisfied, the quasiparticles on inner Fermi surface and that on the outer Fermi surface feel the pair potential with opposite signs,
leading to the emergence of the topological superconductivity characterized by the $\mathbb{Z}_2$ topological index $\nu=1$~\cite{kane_13,zhang_10(2)}.
In the following, we choose $(\Delta_0,\Delta_1)=(-0.5t_n,0.3t_n)$ and $(\Delta_0,\Delta_1)=(-0.5t_n,0.5t_n)$ to describe the phases with $\nu=1$ and $\nu=0$, respectively,
where the topological phase boundary is numerically computed in the SM~\cite{sm}.
In Fig.~\ref{fig:figure4}(a) and Fig.~\ref{fig:figure4}(b), we show $G_{\alpha \beta}$ and $P_{\alpha \beta}$ as a function of the bias voltage, respectively.
The solid (dashed) lines denote $G_{21}$ and $P_{21}$ ($G_{12}$ and $P_{12}$) with $\nu=1$.
Since the HMEM mediates the inter-lead scatterings in both directions,
we find $P_{21}\neq0$ and $P_{12}\neq0$, while $G_{21}=G_{12}=0$ at zero bias voltage.
This result clearly agrees with Eq.~(\ref{eq:linear_res}) and Eq.~(\ref{eq:ns_divergence}).
The dotted line shows the results for $\nu=0$.
Due to the absence of propagating edge modes, we find $G_{\alpha \beta} = P_{\alpha \beta}=0$ for bias voltages below the superconducting gap.
In Fig.~\ref{fig:figure4}(c), we show the noise-to-current ratio $F_{\alpha} = C_{\alpha}/(e\bar{I}_{\alpha})$ with $\nu=1$ as a function of the bias voltage.
In the presence of the HMEM, both $F_1$ and $F_2$ show the divergence at zero bias voltage.
As already shown in Fig.~\ref{fig:figure3}(c), the CMEM causes the divergence in only one direction.
Therefore, the proposed experiment enables a clear distinction between HMEMs and CMEMs.
%----------------------------------------------------------------------------
%----------------------------------------------------------------------------

%****************************************************************************************************************
\section{Discussion and Summary}
%****************************************************************************************************************
A recent experiment on the CoSi$_2$/TiSi$_2$ heterostructure~\cite{lin_21} has observed an anomalous proximity effect~\cite{tanaka_04,tanaka_05(1),asano_07,ikegaya_15,ikegaya_16(1)},
suggesting the realization of spin-triplet superconductivity in this system.
Since the heterostructure exhibits strong Rashba-type spin-orbit coupling,
the noncentrosymmetric ($s+p$)-wave pairing with the dominant $p$-wave component~\cite{fujimoto_07,sigrist_07,sato_09(2)}
is considered the most plausible pairing in the CoSi$_2$/TiSi$_2$ heterostructure~\cite{lin_23}. 
However, the anomalous proximity effect can also be induced by spin-singlet superconductors with asymmetric spin-orbit coupling~\cite{tamura_19,ikegaya_21,lee_25}.
In the SM~\cite{sm}, we show that the noncentrosymmetric ($s+p$)-wave superconductor exhibits the divergent noise-to-current ratio
only when the $p$-wave component is dominant (i.e., when the HMEMs exist).
Therefore, the proposed experiment is useful for determining the pairing symmetry in this system.

Recently, the iron-based superconductor FeTe$_{1-x}$Se$_x$ has attracted much attention as a promising candidate
for a higher-order TS hosting Majorana hinge states~\cite{sarma_19,kim_21,zhang_24}.
Theoretical studies have proposed the emergence of both helical Majorana hinge states~\cite{sarma_19} and chiral Majorana hinge states~\cite{kim_21,zhang_24}.
Tunneling spectroscopy addressing the hinge of FeTe$_{1-x}$Se$_x$ has already been performed experimentally~\cite{burch_19}.
We hope that our proposal will inspire further experiments for the definitive verification of the Majorana hinge states in this compound.

In summary, we study the multi-terminal transport in systems consisting of the two NMs and the TS.
We show analytically that the unpaired nature of the dispersing MEMs generally leads to the absence of time-averaged charge currents in the linear response limit,
while the current fluctuation remains significant.
We numerically reproduce the analytical predictions using the tight-binding models of the topological-insulator-based TSs.
Since the current shot noise is more difficult to measure experimentally than the differential conductance,
observing $G_{\alpha \beta}=0$ at zero bias voltage would be an efficient and practical first step in demonstrating the dispersing MEMs.
Nevertheless, observing the divergence of the noise-to-current ratio provides more definitive evidence for the dispersing MEMs.
Our theory suggests a promising way to unambiguously detect the dispersing MEMs in topological superconductors,
such as CoSi$_2$/TiSi$_2$ heterostructures and iron-based superconductors FeTe$_{1-x}$Se$_x$.

\begin{acknowledgments}
We thank K. Hashimoto, J. Shiogai, T. Arakawa, M. Roppongi, and S. Liu for the fruitful discussions. 
S.I. is supported by a Grant-in-Aid for JSPS Fellows (JSPS KAKENHI Grant No. JP22KJ1507) and a Grant-in-Aid for Early-Career Scientists (JSPS KAKENHI Grant No. JP24K17010).
\end{acknowledgments}

%-----------------------------------------------------------------------------------------------------
% Supplemental Material
%-----------------------------------------------------------------------------------------------------
\clearpage
\onecolumngrid
\begin{center}
  \textbf{\large Supplemental Material for \\ ``Noise-to-current ratio divergence as a fingerprint of dispersing Majorana edge modes''}\\ \vspace{0.3cm}

Leo Katayama$^{1}$, Andreas P. Schnyder$^{2}$, Yasuhiro Asano$^{3}$, and Satoshi Ikegaya$^{1,4}$\\ \vspace{0.1cm}
{\itshape $^{1}$Department of Applied Physics, Nagoya University, Nagoya 464-8603, Japan\\
$^{2}$Max-Planck-Institut f\"ur Festk\"orperforschung, Heisenbergstrasse 1, D-70569 Stuttgart, Germany\\
$^{3}$Department of Applied Physics, Hokkaido University, Sapporo 060-8628, Japan\\
$^{4}$Institute for Advanced Research, Nagoya University, Nagoya 464-8601, Japan}
\date{\today}
\end{center}

%-----------------------------------------------------------------------------------------------------
\section{Topological Properties of Superconducting \\ Topological Insulator Thin-Films}
%-----------------------------------------------------------------------------------------------------
In this section, we discuss the topological properties of the superconducting topological insulator thin-film~\cite{zhang_15,sarma_21},
which is described by the tight-binding Hamiltonian in Eq.~(19) of the main text:
\begin{align}
\begin{split}
&H_s = \frac{1}{2} \sum_{\boldsymbol{k}}\left[C_{\boldsymbol{k}}^{\dagger},C_{-\boldsymbol{k}}^{\mathrm{T}}\right] H_{\boldsymbol{k}}
\left[\begin{array}{cc} C_{\boldsymbol{k}} \\ C_{-\boldsymbol{k}}^{\ast} \end{array} \right],\\
&H_{\boldsymbol{k}} = \left[\begin{array}{cc} h_{\boldsymbol{k}} & \Delta_{\boldsymbol{k}}\\
\Delta_{\boldsymbol{k}}^{\dagger} & -h_{-\boldsymbol{k}}^{\ast} \end{array}\right],\\
&C_{\boldsymbol{k}}=\left[c_{\boldsymbol{k}\uparrow t}, c_{\boldsymbol{k}\downarrow t}, c_{\boldsymbol{k}\uparrow b}, c_{\boldsymbol{k}\downarrow b} \right]^{\mathrm{T}},\\
&h_{\boldsymbol{k}} = h^{\mathrm{Dirac}}_{\boldsymbol{k}} + V,\\
&h^{\mathrm{Dirac}}_{\boldsymbol{k}} = v  \rho_z \otimes  (\sin k_y s_x - \sin k_x s_y) + w_{\boldsymbol{k}}\rho_x,\\
&w_{\boldsymbol{k}} = w(2-\cos k_x - \cos k_y),\\
&V = t \rho_x- (\mu_0 + \mu_{\delta} \rho_z) + (m_0 + m_{\delta} \rho_z) \otimes s_z,\\
&\Delta_{\boldsymbol{k}} = \left\{\Delta_0 + \Delta_1 (\cos k_x + \cos k_y) \right\} (i s_y),
\end{split}
\end{align}
where $c_{\boldsymbol{k}s t}$ ($c_{\boldsymbol{k}s b}$) is an annihilation operator of an electron at the top (bottom) layer with momentum $\boldsymbol{k}$ and spin $s=\uparrow, \downarrow$.
The Pauli matrices $\rho_{\nu}$ and $s_{\nu}$ for $\nu=x,y,z$ denote the layer and spin degrees of freedom, respectively.
The Hamiltonian $h^{\mathrm{Dirac}}_{\boldsymbol{k}}$ describes the Dirac surface states.
The second term $w_{\boldsymbol{k}}$ opens the gaps at the spurious Dirac points at $(k_x,k_y)=(\pm \pi, \pm \pi)$, while leaving the Dirac point at $(k_x,k_y)=(0,0)$ gapless.
The coupling between the top and bottom Dirac surface states is denoted by $t$.
The chemical potential at the top (bottom) layer is given by $\mu_0+\mu_{\delta}$ ($\mu_0-\mu_{\delta}$).
The exchange potential at the top (bottom) layer is represented by $m_0+m_{\delta}$ ($m_0-m_{\delta}$).
The extended $s$-wave pair potential acting on the Dirac surface states is denoted by $\Delta_{\boldsymbol{k}}$.

\subsection{class D}
\begin{figure}[h]
\begin{center}
\includegraphics[width=0.3\textwidth]{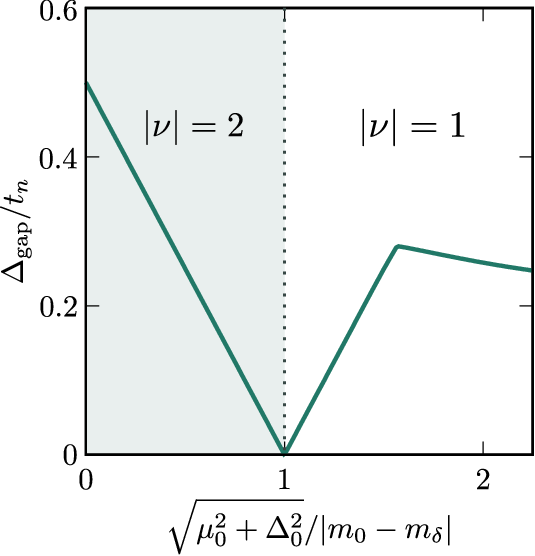}
\caption{Superconducting gap size $\Delta_{\mathrm{gap}}$ as a function of $\sqrt{\mu_0^2+\Delta_0^2}$, where we fix $\mu_0/\Delta_0=1.5$.}
\label{fig:figure_sp1}
\end{center}
\end{figure}

First, we focus on a topological superconductor (TS) belonging to class D, where we set the parameters as $t=\mu_{\delta}=\Delta_1=0$, for simplicity.
By using a unitary operator,
\begin{align}
\begin{split}
& U = U_2 U_1,\\
&U_1 = \left[ \begin{array}{cc}u_1 & 0 \\ 0 & u_1 \end{array} \right],\quad
u_1 = \left[ \begin{array}{cccc}
\alpha & 0 & \beta & 0 \\
0 & \alpha & 0 & -\beta \\
0 & \beta & 0 & \alpha \\
\beta & 0 & -\alpha & 0 \end{array} \right],\\
&\alpha = \frac{A_{\boldsymbol{k}}+m_{\delta}}{\sqrt{2A_{\boldsymbol{k}}(A_{\boldsymbol{k}}+m_{\delta})}}, \quad
\beta = \frac{w_{\boldsymbol{k}}}{\sqrt{2A_{\boldsymbol{k}}(A_{\boldsymbol{k}}+m_{\delta})}}, \quad
A_{\boldsymbol{k}}=\sqrt{w_{\boldsymbol{k}}^2+m_{\delta}^2},\\
&U_2 = \left[ \begin{array}{cccc}
\sigma_0& 0 & 0 & 0 \\
0 & 0 & \sigma_0 & 0 \\
0 & \sigma_0 & 0 & 0 \\
0 & 0 & 0 & \sigma_0 \end{array} \right], \quad
\sigma_0 = \left[ \begin{array}{cc}1 & 0 \\ 0 & 1 \end{array} \right],
\end{split}
\end{align}
we deform the Hamiltonian as
\begin{align}
\begin{split}
&H^{\prime}_{\boldsymbol{k}}=UH_{\boldsymbol{k}}U^{\dagger}=
\left[ \begin{array}{cc} H_{\boldsymbol{k},+} & V_{\boldsymbol{k}} \\ V^{\dagger}_{\boldsymbol{k}} & H_{\boldsymbol{k},-} \end{array} \right],\\
& H_{\boldsymbol{k},\pm}=\left[ \begin{array}{cc}
h_{\boldsymbol{k},\pm} &\Delta_0( m_{\delta}/A_{\boldsymbol{k}}) (i \sigma_y)\\
-\Delta_0( m_{\delta}/A_{\boldsymbol{k}}) (i \sigma_y) & -h^{\ast}_{-\boldsymbol{k},\pm} \end{array} \right],\\
&h_{\boldsymbol{k},\pm}=(A_{\boldsymbol{k}}\pm m_0) \sigma_z -\mu_0 + v(\sin k_y \sigma_x \mp \sin k_x \sigma_y ),\\
&V_{\boldsymbol{k}}= \left[ \begin{array}{cc}
0 &\Delta_0( w_{\boldsymbol{k}}/A_{\boldsymbol{k}}) \sigma_z\\
-\Delta_0( w_{\boldsymbol{k}}/A_{\boldsymbol{k}}) \sigma_z & 0 \end{array} \right],
\end{split}
\end{align}
where $\boldsymbol{\sigma}=(\sigma_x,\sigma_y,\sigma_z)$ are the three Pauli matrices.
When $m_0+m_{\delta}>\mu_0$, the normal bands described by $h_{\boldsymbol{k},+}$ pinch off from the Fermi level and do not affect the low-energy excitation, $E \ll \Delta_0$.
Therefore, in this parameter regime, we obtain the effective Hamiltonian,
\begin{align}
H^{\mathrm{eff}}_{\boldsymbol{k}} = H_{\boldsymbol{k},-} = \left[ \begin{array}{cc}
h_{\boldsymbol{k},-} & \Delta_0( m_{\delta}/A_{\boldsymbol{k}}) (i \sigma_y)\\
-\Delta_0( m_{\delta}/A_{\boldsymbol{k}}) (i \sigma_y) & -h^{\ast}_{-\boldsymbol{k},-} \end{array} \right],
\end{align}
which is equivalent to Eq.~(22) in the main text.
The effective Hamiltonian $H^{\mathrm{eff}}_{\boldsymbol{k}}$ is equivalent to the Hamiltonian of a superconducting quantum anomalous Hall insulator~\cite{zhang_10}.
When $m_0>m_{\delta}$ and $w > m_0(m_0-m_{\delta})$, the energy spectrum becomes gapless at
$\sqrt{\mu_0^2+\Delta_0^2}=m_0-m_{\delta}$, $\sqrt{\mu_0^2+\Delta_0^2}=\sqrt{(2w)^2+m_{\delta}^2}-m_0$, and $\sqrt{\mu_0^2+\Delta_0^2}=\sqrt{(4w)^2+m_{\delta}^2}-m_0$.
Therefore, according to Ref.~[\onlinecite{zhang_10}], the Hamiltonian $H^{\mathrm{eff}}_{\boldsymbol{k}}$ is characterized by
a $\mathbb{Z}$ topological invariant, with a value of $\nu=-2$ for $\sqrt{\mu_0^2+\Delta_0^2}<m_0-m_{\delta}$,
and $\nu=-1$ for $m_0-m_{\delta}< \sqrt{\mu_0^2+\Delta_0^2}<\sqrt{(4w)^2+m_{\delta}^2}-m_0$.
In Fig.~\ref{fig:figure_sp1}, we show the superconducting gap size $\Delta_{\mathrm{gap}}$ as a function of $\sqrt{\mu_0^2+\Delta_0^2}$, where we fix $\mu_0/\Delta_0=1.5$.
The superconducting gap size $\Delta_{\mathrm{gap}}$ (i.e., the minimum positive energy eigenvalue) is obtained by diagonalizing $H_{\boldsymbol{k}}$ numerically.
The parameters are chosen as $v=t_n$, $w=2t_n$, $m_0=t_n$, $m_{\delta}=0.5t_n$.
We confirm that the superconducting gap vanishes at $\sqrt{\mu_0^2+\Delta_0^2} \approx (m_0-m_{\delta})$.
In the main text, we choose $(\mu_0,\Delta_0)=(0.15t_n,0.1t_n)$ and $(\mu_0,\Delta_0)=(0.75t_n,0.5t_n)$ to describe the topological phases with $|\nu|=2$ and $|\nu|=1$, respectively.
%----------------------------------------------------------------------------
%----------------------------------------------------------------------------

\subsection{class DIII}
\begin{figure}[h]
\begin{center}
\includegraphics[width=0.85\textwidth]{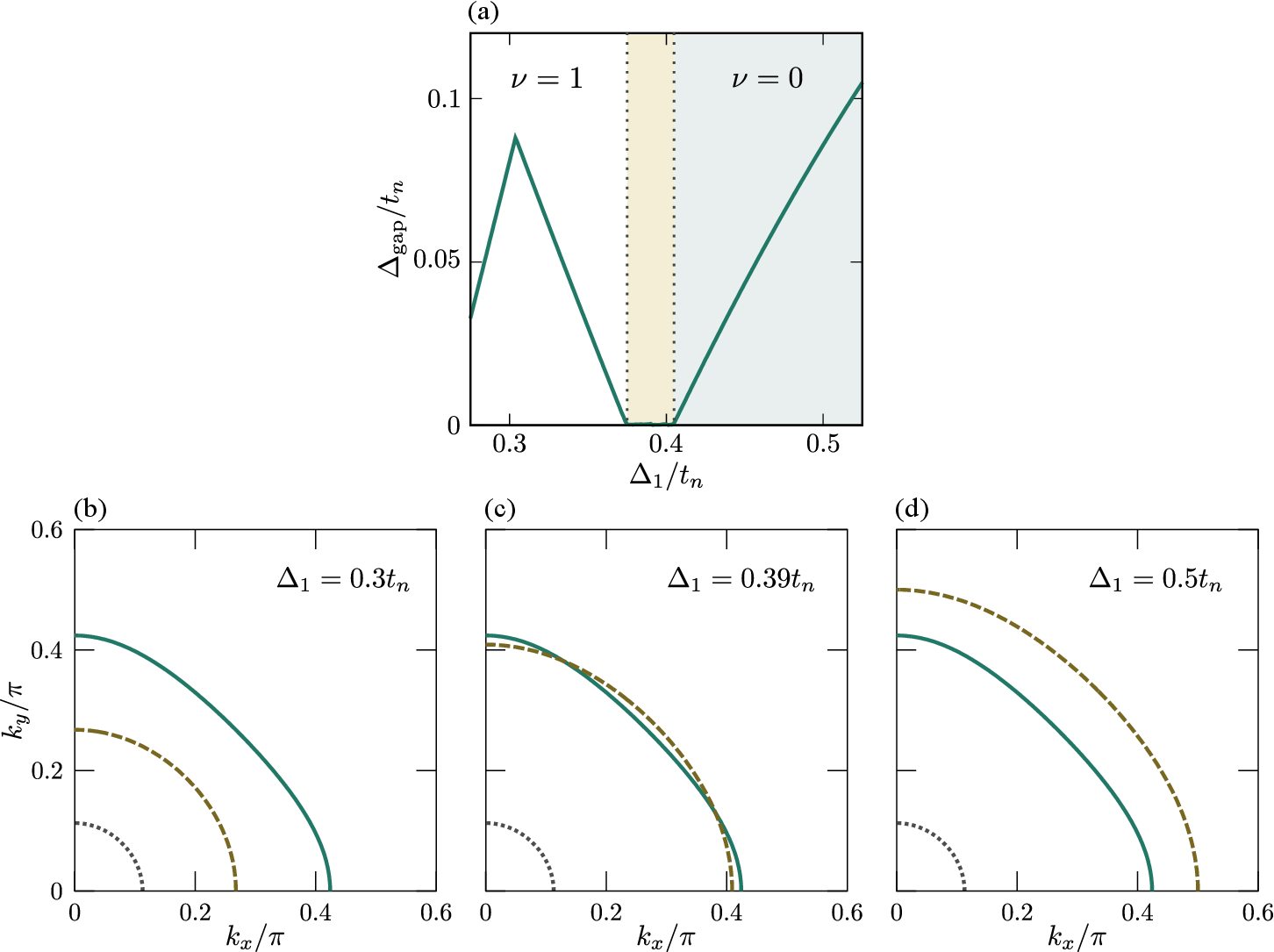}
\caption{(a) Superconducting gap size $\Delta_{\mathrm{gap}}$ as a function of $\Delta_1$.
(b)-(d) Fermi surfaces at $\Delta_1 = 0.3t_n$, $0.39t_n$, and $0.5t_n$, respectively.
The solid (dotted) line denotes the Fermi surfaces of $\boldsymbol{k}_F^+$ ($\boldsymbol{k}_F^-$).
The dashed line represents $\boldsymbol{k}_{\Delta}$.}
\label{fig:figure_sp2}
\end{center}
\end{figure}

Next, we consider the TS belonging to class DIII, where we remove the exchange potential to preserve time-reversal symmetry (i.e., $m_0=m_{\delta}=0$).
By diagonalizing $h_{\boldsymbol{k}}$, we obtain the energy eigenvalues of the normal state as,
\begin{align}
\begin{split}
&\varepsilon_{\boldsymbol{k},\pm}=\sqrt{(\xi_{\boldsymbol{k}}\mp \mu_{\delta})^2+(w_{\boldsymbol{k}} + t)^2}-\mu_0,\\
&\xi_{\boldsymbol{k}}=v\sqrt{\sin^2 k_x + \sin^2 k_y}.
\end{split}
\end{align}
For $t<\mu_0<\sqrt{\mu_{\delta}^2+t^2}$, only $\varepsilon_{\boldsymbol{k},-}$ brunch forms the Fermi surfaces.
For $\sqrt{\mu_{\delta}^2+t^2}<\mu_0<\sqrt{\mu_{\delta}^2+(t+2w)^2}$,
both $\varepsilon_{\boldsymbol{k},+}$ and $\varepsilon_{\boldsymbol{k},-}$ individually form a Fermi surface surrounding the $\Gamma$ point of the Brillouin zone.
Specifically, the Fermi surfaces are located at momenta $\boldsymbol{k}^{\pm}_F$ satisfying $\varepsilon_{\boldsymbol{k}^{\pm}_F,\pm}=0$,
where $|\boldsymbol{k}_F^-| < |\boldsymbol{k}_F^+|$.
The extend $s$-wave pair potential has zeros at $\boldsymbol{k}_{\Delta}=(k_{\Delta,x},k_{\Delta,y})$ satisfying
\begin{align}
\Delta_0 + \Delta_1 (\cos k_{\Delta,x} + \cos k_{\Delta,y}) = 0.
\end{align}
When the condition,
\begin{align}
|\boldsymbol{k}_F^-| < |\boldsymbol{k}_{\Delta}| < |\boldsymbol{k}_F^+|,
\end{align}
is satisfied, the inner Fermi surface and the outer Fermi surface feel the pair potential with opposite signs.
This condition leads to the emergence of the topological superconductivity characterized by $\mathbb{Z}_2$ topological index $\nu=1$~\cite{zhang_10(2),kane_13}.
In Fig.~\ref{fig:figure_sp2}(a), we show the superconducting gap size $\Delta_{\mathrm{gap}}$ as a function of $\Delta_1$,
where we choose $v=t_n$, $w=0.5t_n$, $t=0.5t_n$, $\mu_0=t_n$, $\mu_{\delta}=0.5t_n$, and $\Delta_0 = - 0.5t_n$.
The superconducting gap size $\Delta_{\mathrm{gap}}$ is obtained by diagonalizing $H_{\boldsymbol{k}}$ numerically.
We find that the superconducting gap vanishes around at $0.375t_n < \Delta_1 <0.405t_n$.
For this parameter region, we obtain the nodal superconducting states.
In Fig.~\ref{fig:figure_sp2}(c), we show the Fermi surfaces at $\Delta_1=0.39t_n$,
where the solid and dotted lines denote the Fermi surfaces of $\boldsymbol{k}_F^+$ and $\boldsymbol{k}_F^-$, respectively.
The dashed line represents the momenta where the pair potential vanishes, i.e., $\boldsymbol{k}_{\Delta}$.
The gap nodes appear at momenta where the solid and dashed lines intersect.
For $ \Delta_1 <0.375t_n$, we obtain the topologically nontrivial phase with $\nu=1$,
where the condition of $|\boldsymbol{k}_F^-| < |\boldsymbol{k}_{\Delta}| < |\boldsymbol{k}_F^+|$ is satisfied as shown in Fig.~\ref{fig:figure_sp2}(b).
For $\Delta_1>0.405t_n$, we obtain the topologically trivial phase with $|\boldsymbol{k}_F^-| < |\boldsymbol{k}_F^+| < |\boldsymbol{k}_{\Delta}| $ as shown in Fig.~\ref{fig:figure_sp2}(d).
In the main text,  we choose $(\Delta_0,\Delta_1)=(-0.5t_n,0.3t_n)$ and $(\Delta_0,\Delta_1)=(-0.5t_n,0.5t_n)$ to describe the phases with $\nu=1$ and $\nu=0$, respectively.
%----------------------------------------------------------------------------
%----------------------------------------------------------------------------

%-----------------------------------------------------------------------------------------------------
\section{Noncentrosymmetric Superconductors}
\begin{figure}[h]
\begin{center}
\includegraphics[width=0.6\textwidth]{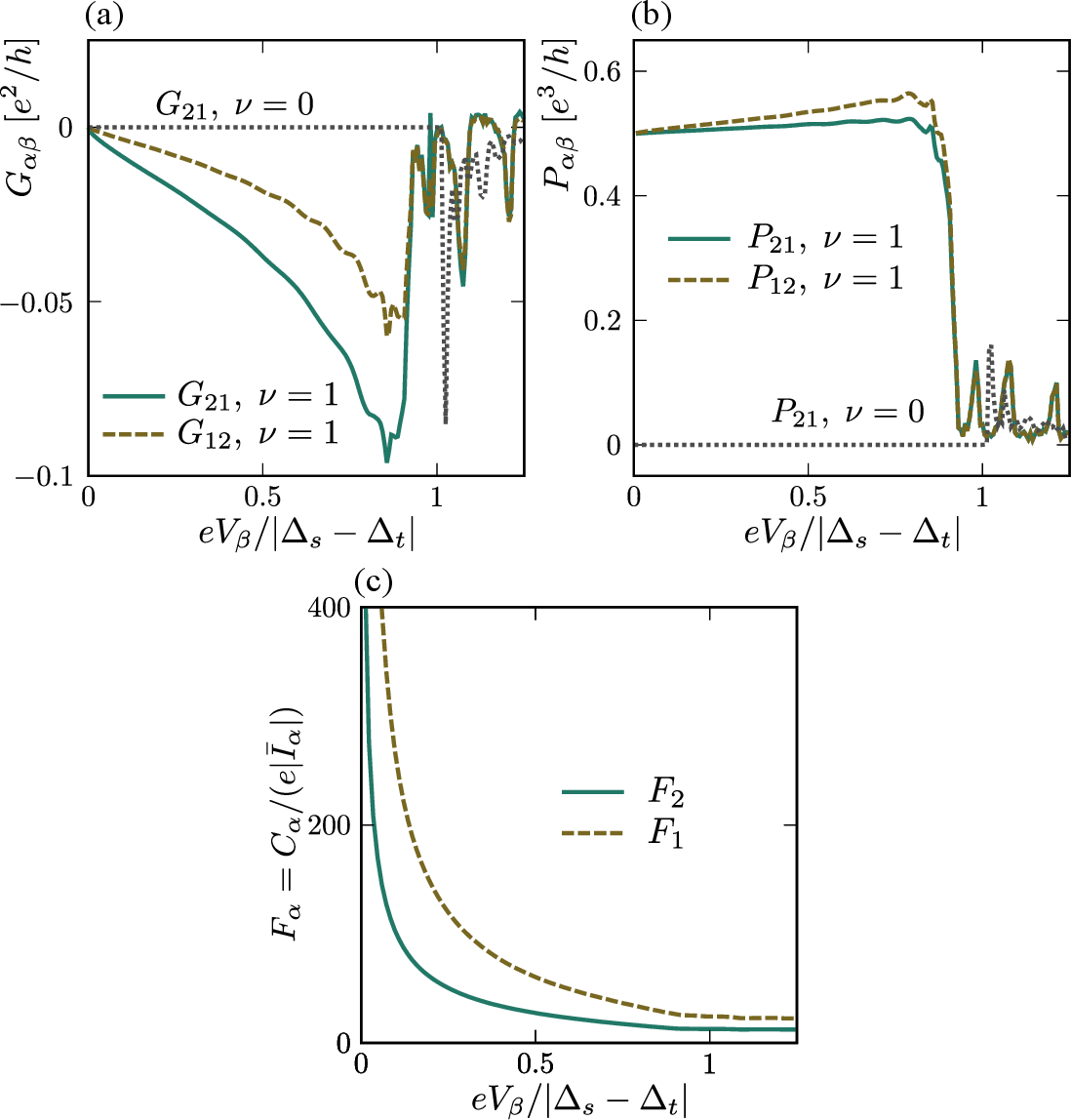}
\caption{(a) Differential conductance $G_{\alpha \beta}$, (b) $P_{\alpha \beta}=dC_{\alpha}/dV_{\beta}$
and (c) noise-to-current ratio $F_{\alpha} = C_{\alpha}/(e\bar{I}_{\alpha})$ for $\nu=1$ as a function of the bias voltage.
The solid (dotted) lines show the results for $\nu=1$ ($\nu=0$) with $(\alpha, \beta)=(2,1)$.
The dashed lines represents the results for $\nu=1$ with $(\alpha, \beta)=(1,2)$.}
\label{fig:figure_sp3}
\end{center}
\end{figure}

%-----------------------------------------------------------------------------------------------------
In this section, we study the transport properties of a noncantrosymmetric ($s+p$)-wave superconductor.
We describe the present superconductor using a two-dimensional tight-binding model on a square lattice, as shown in Fig.~2 of the main text.
The Bogoliubov--de Gennes (BdG) Hamiltonian reads~\cite{fujimoto_07,sigrist_07,sato_09(2)},
\begin{align}
\begin{split}
&H_s = \frac{1}{2} \sum_{\boldsymbol{k}}\left[C_{\boldsymbol{k}}^{\dagger},C_{-\boldsymbol{k}}^{\mathrm{T}}\right] H_{\boldsymbol{k}}
\left[\begin{array}{cc} C_{\boldsymbol{k}} \\ C_{-\boldsymbol{k}}^{\ast} \end{array} \right],\\
&H_{\boldsymbol{k}} = \left[\begin{array}{cc} h_{\boldsymbol{k}} & \Delta_{\boldsymbol{k}}\\
\Delta_{\boldsymbol{k}}^{\dagger} & -h_{-\boldsymbol{k}}^{\ast} \end{array}\right],\\
&C_{\boldsymbol{k}}=\left[c_{\boldsymbol{k}\uparrow}, c_{\boldsymbol{k}\downarrow} \right]^{\mathrm{T}},\\
&h_{\boldsymbol{k}} = \xi_{\boldsymbol{k}}+ \lambda \boldsymbol{g}_{\boldsymbol{k}} \cdot \boldsymbol{s},\\
&\Delta_{\boldsymbol{k}} = \left[ \Delta_s + \Delta_t \boldsymbol{g}_{\boldsymbol{k}} \cdot \boldsymbol{s} \right]\left( i s_y \right),\\
&\xi_{\boldsymbol{k}}=  2t_s(1-\cos k_x) - 2t_s(1-\cos k_y) - \mu_s, \\
& \boldsymbol{g}_{\boldsymbol{k}} = (\sin k_y, -\sin k_x, 0),
\end{split} \label{eq:tb_ham_ncs}
\end{align}
where $c_{\boldsymbol{k}s}$ is an annihilation operator of an electron with momentum $\boldsymbol{k}$ and spin $s=\uparrow, \downarrow$,
$t_s$ is the nearest-neighbor hopping integral, $\mu_s$ represents the chemical potential.
The strength of Rashba spin-orbit coupling is given by $\lambda$.
The pair potential contains both a spin-singlet $s$-wave component $\Delta_s$ and a spin-triplet $p$-wave component $\Delta_t$.
Here we briefly discuss the topological properties of the noncentrosymmetric ($s+p$)-wave superconductor.
The positive energy eigen values of $H_{\boldsymbol{k}}$ are given by
\begin{align} 
E_{\boldsymbol{k},\pm} = \sqrt{(\xi_{\boldsymbol{k}}\pm \lambda |\boldsymbol{g}_{\boldsymbol{k}}|)^2+
(\Delta_s \pm \Delta_t |\boldsymbol{g}_{\boldsymbol{k}}|)^2}.
\end{align}
The superconducting gap vanishes at
\begin{align}
\Delta_s = \Delta_c = \Delta_t |\boldsymbol{g}_{\boldsymbol{k}_F}|,
\end{align}
where $\boldsymbol{k}_F$ satisfies $\xi_{\boldsymbol{k}_F}- \lambda |\boldsymbol{g}_{\boldsymbol{k}_F}|=0$.
For $\Delta_s > \Delta_c$ ($\Delta_s < \Delta_c$),
the BdG Hamiltonian $H_{\boldsymbol{k}}$ can be deformed into
the BdG Hamiltonian of a pure spin-singlet $s$-wave (pure spin-triplet helical $p$-wave) superconductor without any gap closing
by decreasing $\lambda$ and $\Delta_t$ ($\Delta_s$) adiabatically.
The helical $p$-wave superconductor is a time-reversal invariant topological superconductor characterized by a $\mathbb{Z}_2$ topological invariant of $\nu=1$.
Thus, for $\Delta_s > \Delta_c$, the noncentrosymmetric ($s+p$)-wave superconductor is topologically equivalent to the helical $p$-wave superconductor
and hosts the helical Majorana edge mode (HMEM)~\cite{fujimoto_07,sigrist_07,sato_09(2)}.
For the numerical calculations, we rewrite $H_s$ in a real space basis by applying a Fourier transformation: $c_{\boldsymbol{k}s} \rightarrow c_{\boldsymbol{r}s}$.
The $\alpha$-th normal metal is described by
\begin{align}
H_{N_{\alpha}}=-t_n \sum_{\boldsymbol{r}\in N_{\alpha}}\sum_{s=\uparrow,\downarrow}\sum_{\boldsymbol{\nu}=\boldsymbol{x},\boldsymbol{y}}
(d_{\boldsymbol{r}+\boldsymbol{\nu}s}^{\dagger}d_{\boldsymbol{r}s} + \mathrm{H.c.})
+(4t_n-\mu_n) \sum_{\boldsymbol{r}\in N_{\alpha}}\sum_{s=\uparrow,\downarrow}d_{\boldsymbol{r}s}^{\dagger}d_{\boldsymbol{r}s},
\end{align}
where $d_{\boldsymbol{r}s}$ annihilates an electron at a site $\boldsymbol{r}$ with spin $s$,
and $\sum_{\boldsymbol{r}\in N_{\alpha}}$ represents the summation over the lattice sites in the $\alpha$-th normal metal.
We describe the junction interface between the noncentrosymmetric ($s+p$)-wave superconductor and the $\alpha$-th normal metal by
\begin{align}
H_{T_{\alpha}} =-t_{\alpha}\sum_{m>m_{\alpha}}^{M_{\alpha}}\sum_{s=\uparrow,\downarrow}(c_{j=1,m,s}^{\dagger}d_{j=0,m,s}+\mathrm{H.c.})
-t_{\alpha}\sum_{m>m_{\alpha}}^{M_{\alpha}}\sum_{s=\uparrow,\downarrow}(c_{1,m,s}^{\dagger}d_{0,m,s}+\mathrm{H.c.}),
\end{align}
where we assume that the coupling is independent of the spins, while it depends on the leads.
In the following calculations, we fix the parameters as $W_s=200$, $W_n=20$, $L=100$, $m_1=30$, $t_s=\mu_s=\mu_N=t_n$, $\lambda=0.2t_n$, $t_1=0.4t_n$ and $t_2=0.6t_n$.
To describe the topologically nontrivial phase with $\nu=1$ and the topologically trivial phase with $\nu=0$,
we choose $(\Delta_s,\Delta_t)=(0.1t_n,0.2t_n)$ and $(\Delta_s,\Delta_t)=(0.2t_n,0.1t_n)$, respectively.
In Fig.~\ref{fig:figure_sp3}(a) and Fig.~\ref{fig:figure_sp3}(b), we show $G_{\alpha \beta}$ and $P_{\alpha \beta}$ as a function of the bias voltage, respectively.
The dotted line shows the results for $\nu=0$.
Due to the absence of propagating edge modes, we find $G_{\alpha \beta} = P_{\alpha \beta}=0$ for bias voltages below the superconducting gap.
The solid (dashed) lines denote $G_{21}$ and $P_{21}$ ($G_{12}$ and $P_{12}$) with $\nu=1$.
Since the HMEM mediates the inter-lead scatterings in both directions,
we find $P_{21}\neq0$ and $P_{12}\neq0$, while $G_{21}=G_{12}=0$ at zero bias voltage.
Therefore, as demonstrated in Fig.~\ref{fig:figure_sp3}(c), we obtain the divergence of the noise-to-current ratio at zero bias voltage.
The divergent noise-to-current ratio is a definitive signature of the dominant $p$-wave pairing that is responsible for the appearance of the HMEM.
%----------------------------------------------------------------------------
%----------------------------------------------------------------------------

\end{document}